\documentclass[apjl]{emulateapj}
\usepackage{color}

\def\gsim{\mathrel{\rlap{\lower 4pt \hbox{\hskip 1pt $\sim$}}\raise 1pt \hbox {$>$}}} \def\lsim{\mathrel{\rlap{\lower 4pt \hbox{\hskip 1pt $\sim$}}\raise 1pt \hbox {$<$}}}

\shorttitle{Rise Time of Type Ia SN 2012ht} \shortauthors{Yamanaka et al.}

\begin{document}

\title{Early-phase photometry and spectroscopy of transitional Type Ia SN 2012ht: 
 Direct constraint on the rise time}

\author{Masayuki \textsc{Yamanaka}\altaffilmark{1},
        Keiichi \textsc{Maeda}\altaffilmark{2}, 
        Miho \textsc{Kawabata}\altaffilmark{3},    
        Masaomi \textsc{Tanaka}\altaffilmark{4},   
        Katsutoshi \textsc{Takaki}\altaffilmark{5},
        Issei \textsc{Ueno}\altaffilmark{5}, 
        Kazunari \textsc{Masumoto}\altaffilmark{3},    
        Koji S. \textsc{Kawabata}\altaffilmark{6},	
        Ryosuke \textsc{Itoh}\altaffilmark{5},
	 Yuki \textsc{Moritani}\altaffilmark{6},
        Hiroshi \textsc{Akitaya}\altaffilmark{6}, 
        Akira \textsc{Arai}\altaffilmark{7},
        Satoshi \textsc{Honda}\altaffilmark{7}, 
        Koichi \textsc{Nishiyama}\altaffilmark{8},
        Fujio \textsc{Kabashima}\altaffilmark{9},        
        Katsura \textsc{Matsumoto}\altaffilmark{3},
	 Daisaku \textsc{Nogami}\altaffilmark{1}, and
	 Michitoshi \textsc{Yoshida}\altaffilmark{6}
}

\altaffiltext{1}{Kwasan Observatory, Kyoto University, 
17-1 Kitakazan-ohmine-cho, Yamashina-ku, Kyoto, 607-8471, Japan; yamanaka@kwasan.kyoto-u.ac.jp} 
\altaffiltext{2}{Department of Astronomy, Kyoto University, Kitashirakawa-Oiwake-cho, Sakyo-ku, Kyoto 606-8502, Japan}
\altaffiltext{3}{Astronomical Institute, Osaka Kyoiku University, Asahigaoka, Kashiwara, Osaka 582-8582, Japan}
\altaffiltext{4}{National Astronomical Observatory of Japan, Osawa, Mitaka, Tokyo 181-8588, Japan}
\altaffiltext{5}{Department of Physical Science, Hiroshima University, Kagamiyama 1-3-1, Higashi-Hiroshima 739-8526, Japan} 
\altaffiltext{6}{Hiroshima Astrophysical Science Center, Hiroshima University, Higashi-Hiroshima, Hiroshima 739-8526, Japan}
\altaffiltext{7}{Center for Astronomy, University of Hyogo, 407-2 Nishigaichi, Sayo-cho, Sayo, Hyogo 679-5313}
\altaffiltext{8}{Kurume, Fukuoka-ken, Japan}
\altaffiltext{9}{Miyaki-cho, Saga-ken, Japan}



\begin{abstract}

 We report photometric and spectroscopic observations of 
 the nearby Type Ia Supernova (SN Ia) 2012ht from $-15.8$ days 
 to $+49.1$ days after $B$-band maximum. The decline rate of the 
 light curve is $\Delta m_{15}$($B$)$=1.39~\pm~0.05$ mag, 
 which is intermediate between normal and subluminous SNe Ia,
 and similar to that of the `transitional' Type Ia SN 2004eo.
  The spectral line profiles also 
  closely resemble those of SN 2004eo.
We were able to observe SN 2012ht at very early phase,
when it was still rising and was about three magnitudes 
 fainter than at the peak. 
The rise time to the $B$-band maximum is estimated 
 to be $17.6 \pm 0.5$ days 
 and the time of the explosion is MJD $56277.98 \pm 0.13$. 
 SN 2012ht is the first transitional SN Ia whose rise time is directly
   measured without using light curve templates, 
   and the fifth SN Ia overall.
 This rise time is consistent with 
 those of the other four SNe within
 the measurement error, even including the extremely early detection of 
 SN 2013dy.
 The rising part of the light curve can be fitted by a quadratic function,
 and shows no sign of a shock-heating component due to
 the interaction of the ejecta with a companion star. 
 The rise time is significantly longer than that inferred for 
 subluminous SNe such as SN 1991bg, which
 suggests that a progenitor and/or explosion mechanism of 
 transitional SNe Ia are more similar to normal SNe Ia rather than
 subluminous SNe Ia.

\end{abstract}


\section{Introduction}

 When the mass of a carbon-oxygen white dwarf approaches the 
 Chandrasekhar limiting mass, a thermonuclear 
 runaway starts, 
 and the whole star explodes. This is thought to be the origin
 of Type Ia supernova (SN Ia).  
 Since the explosion occurs with this limiting mass,
 observational properties of SNe Ia are homogeneous.
 The peak luminosity is correlated with the decline rate of the light curves 
 \citep{Phillips1993}. Using these properties, the acceleration of the 
 cosmic expansion was discovered \citep{Perlmutter1999, Riess1998}.
 In spite of the importance of SNe Ia in the field of astrophysics 
 and cosmology, the progenitor scenario and explosion mechanism 
 are not yet well understood
 \citep{Hillebrandt2000}.

 The rising part of SNe Ia can provide important constraints 
 on the progenitor system \citep[e.g., ][]{Kasen2010}
 and the explosion mechanism \citep[e.g., ][]{Piro2010}.
 Recently, detailed studies on the 
 very early phases have been reported for
 several objects, which include
 SNe 2009ig \citep{Foley2012,Marion2013}, 2011fe \citep{Nugent2011,Bloom2012,Vinko2012,Pereira2013},
 2012cg \citep{Silverman2012a,Munari2013} and 2013dy \citep{Zheng2013}. 
 These SNe are thought to be discovered
   within 2 days after the explosion, 
 at 4 magnitudes fainter than its peak.
 The best case is SN 2011fe.
 The rising part of the light curve was 
 well fitted by a quadratic function 
 from 0.2 days after the explosion time \citep{Bloom2012}. 
 Detailed study of the rising part of 
 the light curves is still limited to a small number of objects.

 In this Letter, we present another excellent observational example of 
a SN Ia at very early phase. We discovered SN 2012ht on 2012 
Dec 18.772 UT at unfiltered
magnitude of $17.15 \pm 0.09$ mag \footnote{The magnitudes on
Dec 18.772 and Dec 19.72 were recalibrated (see \S 2).}
in the nearby galaxy NGC 3447 \citep{NishiyamaKabashima2012},
whose redshift is 0.003559 ($NED$).
An additional observation was performed, showing that the object
brightened to an unfiltered-magnitude of $16.23 \pm 0.08$ mag on
Dec 19.72. On the other hand, this SN was not detected
($\geq 18.6$ mag) on Dec. 16.736 at 2.0 days before the discovery
\citep{NishiyamaKabashima2012}. Spectroscopic observations 
soon confirmed that it was a
young SN Ia \citep{Milisavljevic2012,Yamanaka2012}.
We performed 
photometric and spectroscopic observations of
SN 2012ht from -15.8 days to +49.1 days after the B-band maximum.
As shown below, we determine that the epoch of the discovery is
just 1.8 days after the explosion. We classify this SN as a
transitional SN Ia between normal and subluminous SNe Ia,
based on the light curves and spectral line profiles.
The rise time of SN 2012ht is compared with those of a few
other SNe Ia of which the rise time have been accurately estimated.
We also discuss the absence of excess emission in the earliest
phases and give an implication for a progenitor scenario and
explosion mechanism.

\section{Observations and Data Reduction}

We performed $B$, $V$, $R$ and $I$-band imaging observations 
of SN 2012ht on 33 nights from 2012 Dec 20.7 UT ($-13.9$ days) to 2013
Feb 21.7 ($+49.1$ days) using HOWPol \citep[Hiroshima One-shot Wide-field 
Polarimeter; ][]{Kawabata2008} mounted on the 1.5-m Kanata 
telescope at Higashi-Hiroshima Observatory, and 
on 40 nights from $-10.5$ to $+49.1$ days using the 51cm telescope
in Osaka Kyoiku University.
The images were reduced according to standard procedures of 
CCD photometry. We performed aperture photometry on 
HOWPol data and point-spread-function (PSF) fitting photometry
on the 51cm telescope's
data using $IRAF$.
The magnitude is calibrated with photometric standard stars in Landolt fields  
(Landolt 1992) observed on photometric nights.
We corrected the magnitudes obtained by 
these two instruments using their color terms. 
Color terms were derived through the observations of the 
standard field  M67 \citep{Chevalier1991} on a photometric night. 
The magnitudes obtained with these instruments
are consistent within a systematic error of 0.1-0.2 magnitude at most.
 We also re-reduced the unfiltered discovery image obtained by 
 \cite{NishiyamaKabashima2012} using PSF photometry.
The same applies to the additional image on Dec 19.7.
The magnitudes are calibrated using local standard stars 
 whose $R$ magnitudes are estimated by HOWPol.

For spectroscopy, we obtained optical spectra using
HOWPol on 12 nights from -13.9 to
+7.1 days and also using MALLS installed to the 2.0 m
Nayuta telescope at Nishi-Harima Astronomical Observatory
(NHAO) on four nights from -9.9 to +7.9 days.
Their wavelength coverages and spectral resolutions are
4500--9000 \AA\ and $R=\lambda /\Delta\lambda = 400$
(at 6000 \AA) for HOWPol and 4200--6800 \AA\ and
$R=1200$ (at 6000 \AA) for MALLS.
For the MALLS data on Dec. 26, they are instead
4200--9000 \AA\ and $R=500$ (at 6000 \AA).
We performed wavelength calibration using telluric
emission lines of the sky superimposed on each object
frame for the HOWPol data and using comparison lamp (FeArNe)
frames for the MALLS data. The fluxes have been calibrated
using spectrophotometric standard star data.

 \section{Results}

 \subsection{Light Curves and Color}

\begin{figure*}
  \begin{center}
    \begin{tabular}{c}
 \resizebox{90mm}{!}{\includegraphics{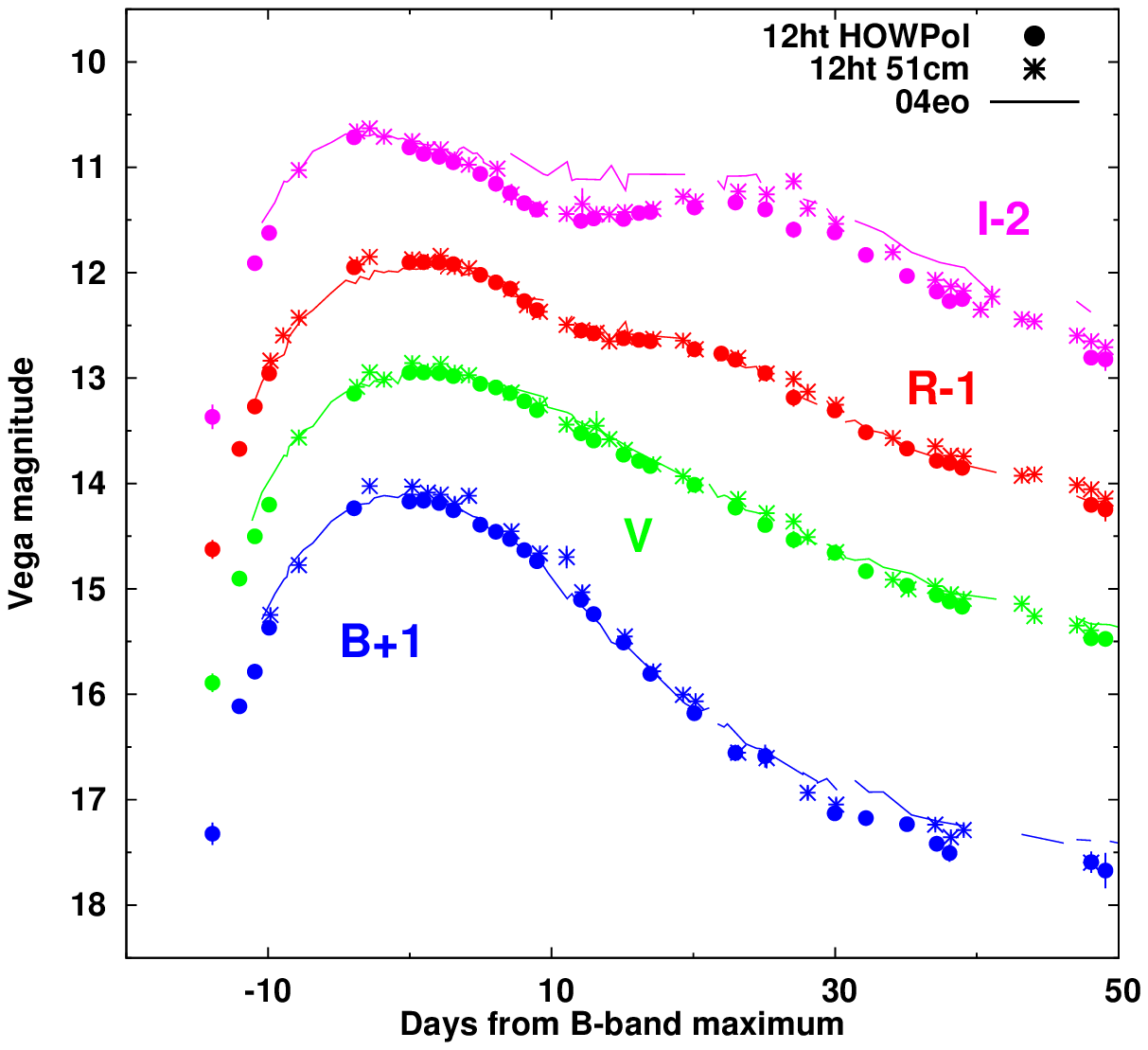}} 
 \resizebox{90mm}{!}{\includegraphics{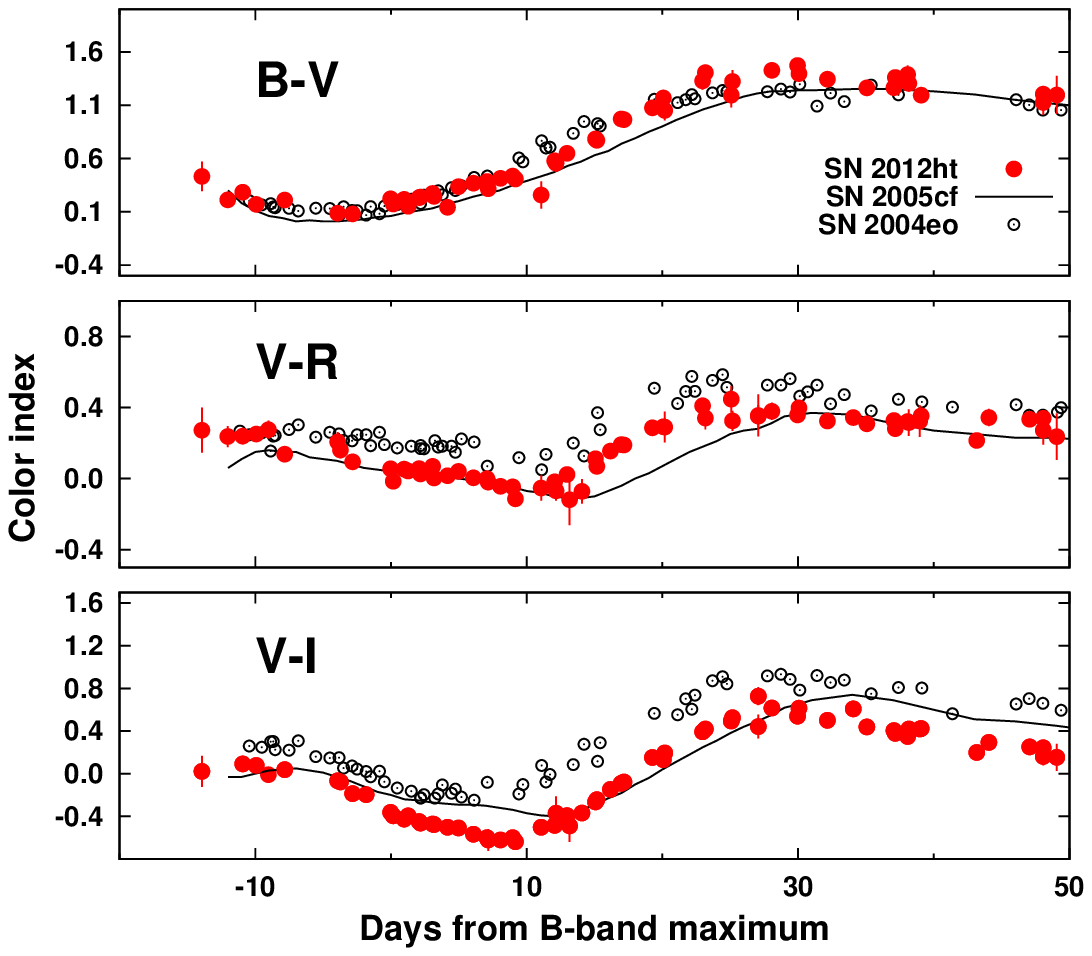}} \\
    \end{tabular}
    \caption{(Left panel) $B$, $V$, $R$ and $I$-band light curves 
      of SN 2012ht compared with those of another transitional SN Ia 2004eo 
      \citep{Pastorello2007b}. Galactic extinction has been corrected for using the values
      from \cite{Schlafly2011}. 
      The filled and open circles denote the data obtained by 1.5m Kanata and 
      51cm telescopes, respectively. 
      The lines denote the light curves of SN 2004eo. 
      The $B$-band maximum date and magnitude are determined as MJD $56295.60 \pm 0.61$ 
      and $13.22 \pm 0.04$ mag by 
      polynomial fitting. The maximum magnitudes of SN 2004eo are shifted 
      to match to those of SN 2012ht for the comparison. 
     (Right panel) $B-V$, $V-R$ and $V-I$ color evolution compared to those of SN 2004eo 
     \citep{Pastorello2007b} and a typical SN Ia 2005cf \citep{XWang2009}. 
     The error bar of each data point denotes the square root of the sum of observational 
     error ($1 \sigma$) and the systematic error of our photometry.
      }
    \label{lc}
  \end{center}
\end{figure*}

 The $B$, $V$, $R$ and $I$-band light curves of SN 2012ht are shown in Figure \ref{lc}. 
 These data have been corrected for the Galactic extinction of $E(B-V)=0.02$ mag 
 and $R_{V}=3.1$ \citep{Schlafly2011}. The extinction within the host galaxy is 
 assumed to be negligible as discussed in the next paragraph.
 The $B$-band maximum date and
 magnitude are derived to be MJD $56295.60 \pm 0.61$ (2013 Jan 3.6 UT) and $13.22 \pm 
 0.04$ mag, respectively, by polynomial fitting at around maximum. 
 Throughout this Letter, the phase $t$ is expressed relative
 to the $B$-band maximum date, i.e., $t = 0$ is MJD 56295.60.
 
 We also use this definition in discussing about the rise time in \S 3.2.
 The light curves 
 of SN 2012ht are similar to those of 
   SN 2004eo \citep[$\Delta m_{15}$($B$)$=1.46$ mag; ][]{Pastorello2007b} 
   in all the bands.
 SN 2004eo is classified as a transitional SN Ia, which is intermediate
 between normal and subluminous SNe Ia \citep{Pastorello2007b}.
 We derived the decline-rate parameter of SN2012ht
 as $\Delta m_{15}$ ($B$)$=1.39 \pm 0.05$ mag, 
 which is close to that of SN 2004eo.
 The time of the $I$ secondary maximum is at $t=22.5 \pm0.5$ day, 
 which is also consistent with that of SN 2004eo.
   A small difference can be seen only 
   in the $I$-band magnitude after the first maximum; 
   SN 2012ht shows a clearer dip between the first
 and secondary maximum than SN 2004eo.
  We confirm the 3.2, 2.9, 2.7 and 2.6 magnitude brightening 
 in $B$, $V$, $R$ and $I$-band, respectively, 
 from $t=-13.9$ to $B$-band maximum. It is remarkable that we performed 
 multi-band photometry at such early phases, 
 thanks to the follow-up observations from a very early phase.




 The distance modulus for NGC 3447 has relatively large uncertainty 
  due to its infall velocity to Virgo cluster \citep{Marino2010}. 
  Adopting the Tully-Fisher distance of $\mu = 31.5 \pm 0.4$ 
  from $NED$, the absolute magnitude of SN 2012ht is calculated as 
  $M_{B}=-18.3 \pm 0.4$ mag. It is noted that its magnitude is fainter than
  $M_{B}=-19.14 \pm 0.10$ mag expected from its decline-rate parameter
  of $\Delta m_{15}$($B$)$=1.39$ mag \citep{Prieto2006}.

 The color evolution of SN 2012ht is shown in the right panel of 
 Figure  \ref{lc}, compared to the intrinsic color ({\it i.e.},
 extinction-corrected color) evolution of SNe 2005cf 
 \citep{XWang2009} and 2004eo \citep{Pastorello2007b}. 
 The $B-V$ color of SN 2012ht is comparable to
 that of SNe 2005cf and 2004eo around maximum.
 This overall similarity of the color evolution suggests
 the extinction in the host galaxy of SN 2012ht is negligible
 based on the Lira relation \citep{Lira1998}. 
 The color evolution in different bandpasses is similar to SN 2004eo, 
 which is of the same subtype as SN 2012ht, except bluer in $V-R$ 
 and $V-I$. This might be a characteristic of SNe Ia with detected carbon 
 absorption \citep[e.g., ][]{Thomas2011}.

 \subsection{Spectral Properties}

\begin{figure*}
 \begin{center}
 \begin{tabular}{c}
\includegraphics[scale=0.8]{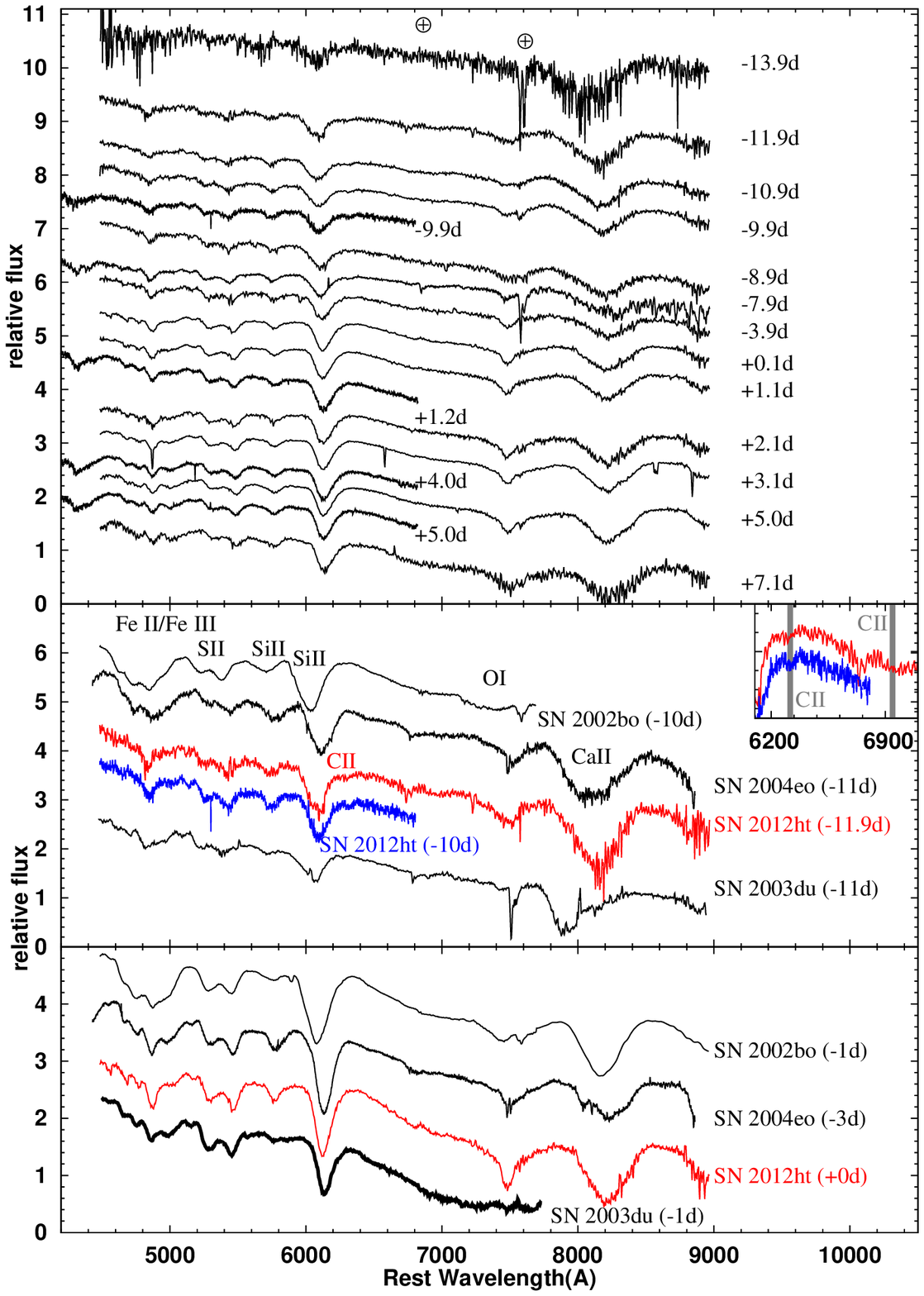}
 \end{tabular}
 \caption{(Upper panel) Time evolution of spectra of SN 2012ht. These spectra are corrected for 
 its recession velocity. Telluric absorption lines remain in the spectrum at -13.9 day due to 
 a low S/N ratio. The phases refer the $B$-band maximum dates at MJD 
 $56295.60 \pm 0.61$. (middle panel) Spectra at $t=-10$ and $-12$ days compared
 with those of SNe 2002bo \citep{Benetti2004}, 2003du \citep{Stanishev2007} and 
 2004eo \citep{Pastorello2007b} at similar phases. The redshifts of host galaxies were
 corrected for. 
 The blue and red lines denote spectra obtained by Nayuta and Kanata telescopes, respectively.
 The tiny right top panel exhibits the close-up of the C~{\sc ii}~$\lambda\lambda$6580,7234 
 absorption lines. The two vertical lines denote these absorption features blueshifted by 
 12000 km~s$^{-1}$.
 (bottom panel) Same as the middle panel at $t=0$.}
  \label{sp}
 \end{center}
 \end{figure*}

  We show the spectral evolution of SN Ia 2012ht from $t=-13.9$ to $+7.1$ days in
 Figure \ref{sp}. Spectra have been corrected for the recession velocity of the host galaxy NGC
 3447 from $NED$. The atmospheric absorption lines have been removed using standard star data
 obtained at the same nights except for the spectrum at
 $t=-13.9$ day. 

  The O~{\sc i}~$\lambda$ 7774 absorption line in SN 2012ht 
 is very strong compared with other SNe Ia 
 at similar epochs as shown in the 
lower two panels in Figure \ref{sp}.
 The equivalent width is measured as 105 \AA~using the spectrum at $t=+0.1d$.
 \cite{Hachinger2006} investigate the relationship 
of the equivalent widths of several
 absorption lines around maximum with the 
decline-rate ($\Delta m_{15}$($B$)) 
 in SNe Ia. The equivalent width of the O~{\sc i}~$\lambda$7774 absorption line 
 implies $\Delta m_{15}$($B$)$=1.4-1.8$ mag.
 This is consistent with the intermediate 
decline rate of 
$\Delta m_{15}$($B$)$=1.4$ mag found for SN 2012ht. 
 The line-depth ratio ($R$(Si~{\sc ii})) 
 of Si~{\sc ii}~$\lambda$5972 to 
 Si~{\sc ii}$\lambda$6355 is 
 an indicator of the absolute magnitude of SNe Ia \citep{Nugent1995}. 
 We measured the ratio as $R$(Si~{\sc ii})$\sim0.34$ using the spectrum 
 at $t=+0.1$ day.
 This value is also consistent with that expected 
 from the relation between $\Delta m_{15}$($B$) and $R$(Si~{\sc ii}) 
 (Figure 14 of \citealt{Blondin2012}).
 This indicates that SN~2012ht is located in the 
 faint end of normal SNe Ia. At 4500 $-$ 5100 \AA, the broad features are blends of 
 Fe~{\sc ii} multiplet (4923, 5018, 5169), Fe~{\sc iii}, and
 Si~{\sc ii}~$\lambda$5056 absorption lines, which are seen in normal and 
 overluminous SNe Ia 
 \citep{Branch2006}. 
 For SN 2012ht, the absorption line at 4850 \AA~is
 very strong after $t=-11.9$ day. 
 It is very similar to the transitional SN Ia 
 2004eo \citep[see the lower two panels in Figure \ref{sp} and][]{Pastorello2007b}. 
 In other transitional SNe Ia 1986G,
 1992A, 2009an, this absorption line is also clearly stronger than 
 those of normal SNe Ia 
 \citep{Phillips1987, Kirshner1993, Sahu2013}. 
 These spectral properties support that SN 2012ht belongs to the transitional group
 like SN 2004eo.

  The shallow C~{\sc ii} $\lambda$ 6580 absorption line is seen around
 6300 \AA\ in spectra until $-8.9$ day, and it disappears by
 $-3.9$ day. We attempt to seek C~{\sc ii}$~\lambda7234$ in our spectra. 
 However, it was not significantly detected in our spectra. 
 The line velocity of C~{\sc ii} in SN 2012ht is measured to be 12000 km~s$^{-1}$,
 which is the same as that of S~{\sc ii}~$\lambda$ 6355 within the 
 measurement error at $t=-11.9$ d. 
 On the other hand, C~{\sc ii} absorption lines are not seen in SN 2004eo \citep{Pastorello2007b}.

 \section{discussion}

\subsection{Estimate of the Rise Time}
 
\begin{figure*}
  \begin{center}
    \begin{tabular}{c}
\resizebox{90mm}{!}{\includegraphics{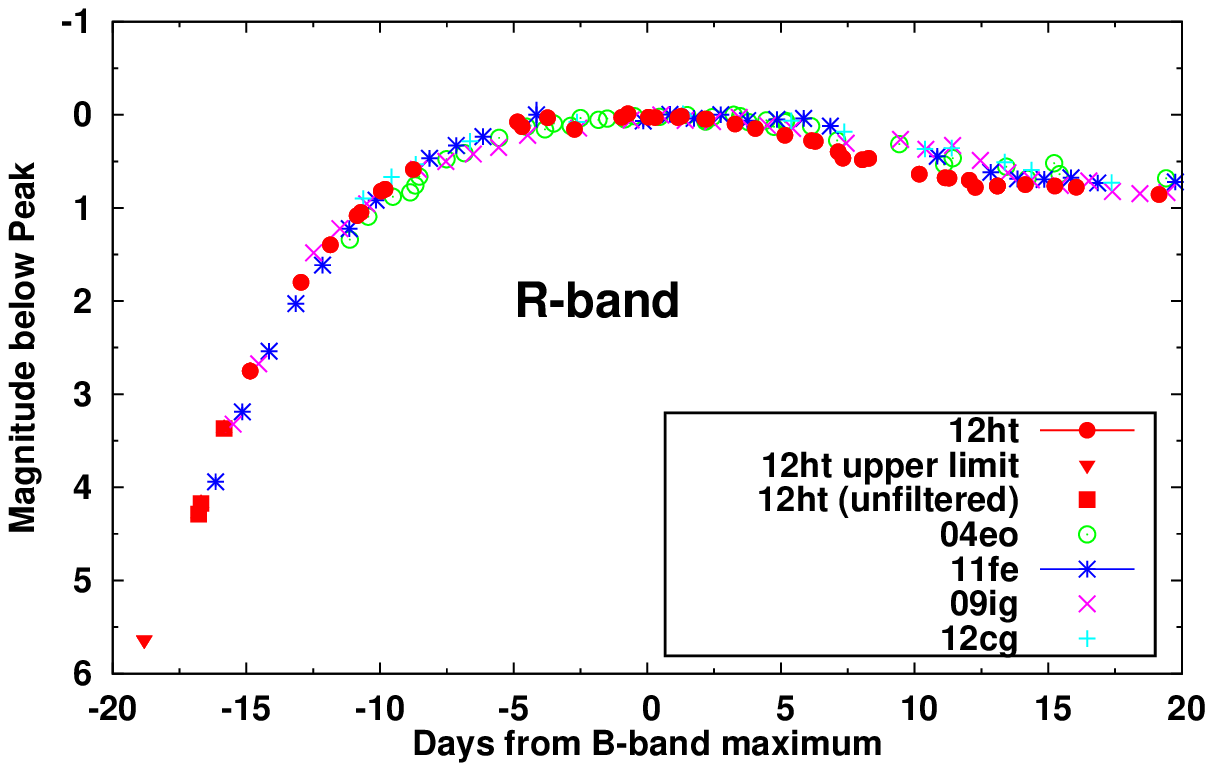}} 
\resizebox{90mm}{!}{\includegraphics{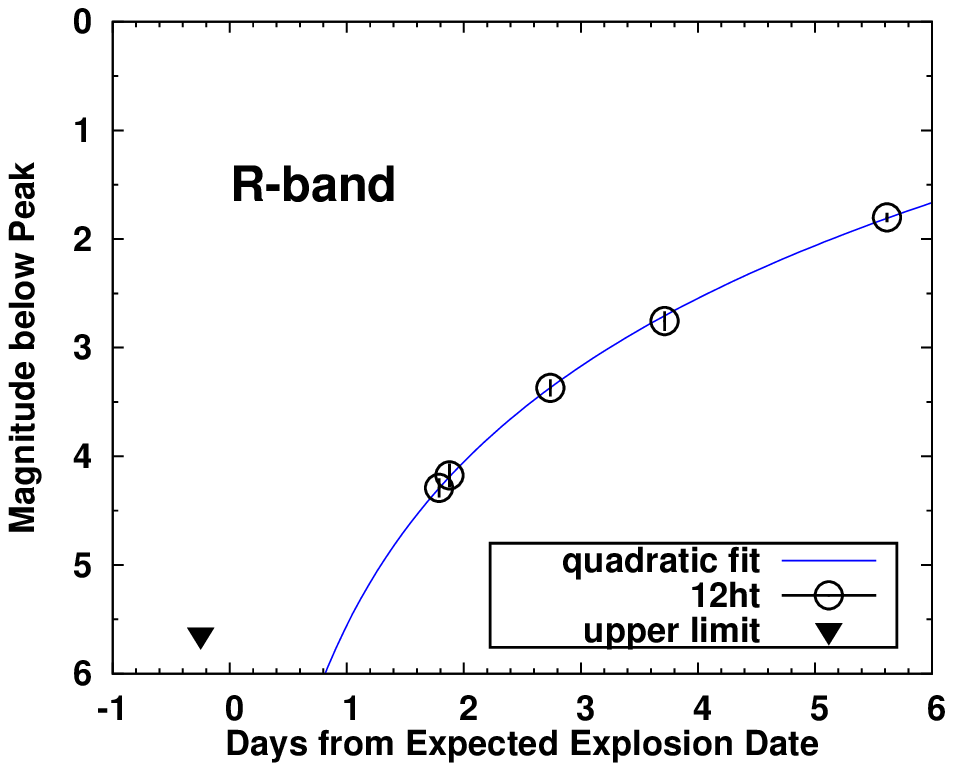}} \\
    \end{tabular}
    \caption{(Left panel) The early part of the $R$-band of SN 2012ht compared with those of 
     SNe 2004eo \citep{Pastorello2007b}, 2009ig \citep{Foley2012} and 2011fe \citep{Vinko2012}. 
      The red filled circles denote the data 
      obtained by 1.5m Kanata and 51cm telescopes. 
      The red filled squares denote the data obtained by Nishiyama and Kabashima.
     The upside down triangle indicates the upper-limit magnitude by Itagaki, K. 
       \citep{NishiyamaKabashima2012}.
       The discovery data have been re-calibrated using $DAOPHOT$.
      (right panel) Close-up of the rising part of the $R$-band light 
      curve since the expected explosion date 
     until $6$ days. 
      The explosion date is estimated as MJD $56277.98 \pm 0.13$ 
      using a fit of a quadratic function denoted by the blue line.
        }
    \label{rt}
  \end{center}
\end{figure*}

  SN 2012ht is the first transitional SN Ia for which the 
 rise time can be accurately estimated by early-phase 
 observations. To give a direct constraint on a rise time from observational data,
 we attempt to fit a quadratic curve to the rising part in the 
 $R$-band light curve of SN 2012ht (see right panel of 
 Figure \ref{rt}). We find 
 that the rising curve up to $t=-8$ day is 
 well fitted within the photometric error. 
 The explosion time was determined as MJD $56277.98 \pm 0.13$, by extending 
 the fitting curve to 
 the zero-point flux level. Surprisingly, the discovery date by K. Nishiyama \& F. Kabashima 
 was $1.8$ days after this explosion time \citep{NishiyamaKabashima2012}, 
 and the upper-limit by K. Itagaki \citep{NishiyamaKabashima2012} was obtained $0.2$ days before the explosion.

\begin{figure*}
  \begin{center}
    \begin{tabular}{c}
 \resizebox{90mm}{!}{\includegraphics{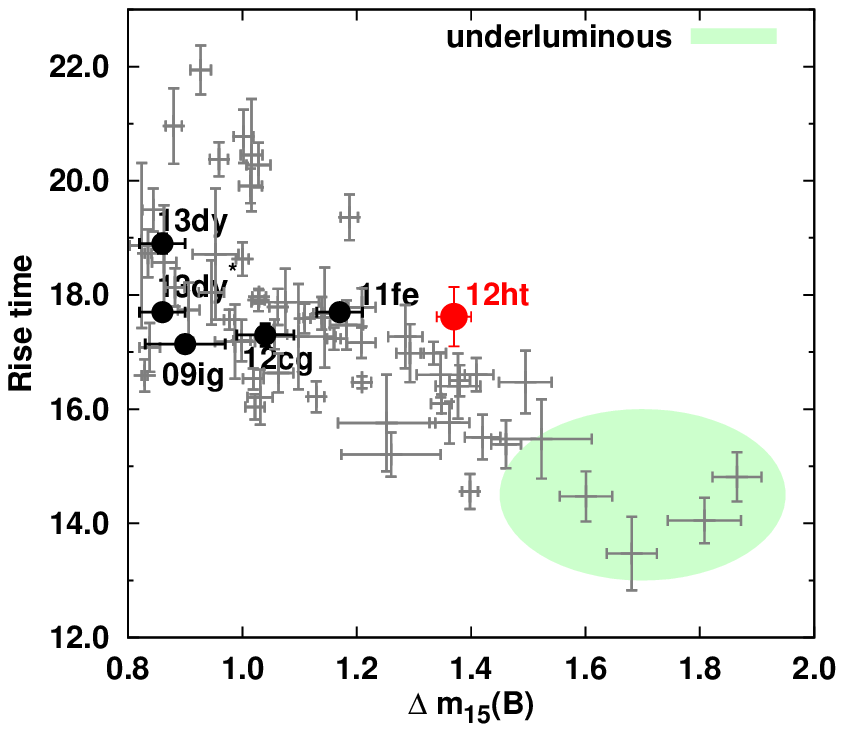}} 
    \end{tabular}
    \caption{The decline rate $\Delta$m$_{15}$($B$) and rise time of SN 2012ht 
  compared with those of SNe 2009ig \citep{Foley2012, Marion2013}, 2011fe \citep{Nugent2011, Pereira2013}, 2012cg \citep{Silverman2012a, Munari2013}, 
  and 2013dy \citep{Zheng2013}. $\Delta$m$_{15}$($B$) of SN 2013dy is 
  0.86 $\pm$ 0.04 mag (W. Zheng, priv. comm.)
  The rise time of SN 2013dy was calculated as 
  17.7 day from the broken power-law fitting, while it is $\sim18.9$ 
 day when fitting with a quadratic function \citep{Zheng2013}. 
 The gray cross points denote the template-fitting data from 
 \cite{Ganeshalingam2011}. The green-shaded area denotes the region that 
 underluminous SNe Ia could occupy.}
    \label{rtc}
  \end{center}
\end{figure*}

  We estimate the rise time of SN 2012ht as $17.62 \pm 0.52$ day\footnote{The 
 rise time is defined as the $B$-band maximum date
 minus the explosion date, being the same definition as in the previous studies
 \citep{Foley2012,Nugent2011,Silverman2012a}}, which
 is compared to those of SNe 2004eo \citep{Pastorello2007b}, 2009ig \citep{Foley2012}, 
 2011fe \citep{Vinko2012} and 2012cg \citep{Munari2013} in the left panel 
 of Figure \ref{rt}. 
 The maximum magnitude is shifted to zero for all the objects. 
 The time is referred to the $B$-band maximum date for each event.
 The rising rate of SN 2012ht is similar to those of 
 SNe 2009ig and 2011fe (Figure \ref{rt}) despite a variety 
 in the luminosity classes covered by these objects (see also Figure \ref{rtc}). 
 SN 2011fe is classified as a very normal SN Ia 
 based on photometric and spectroscopic observations \citep{Vinko2012,Pereira2013}.
 On the other hand, SN 2009ig shows an extremely slow-decline rate in 
 the $B$-band light curve despite the typical absolute magnitude 
 \citep{Foley2012,Marion2013}.  

 We then compare the rise time of SN 2012ht with those of underluminous objects.
Lacking a direct constraint on the rise time of these faint SNe Ia, we rely on  
 statistical studies \citep{Riess1999,Hayden2010,
 Ganeshalingam2011,Gonzalez2012}. 
 To estimate the rise time, they used a 
 template light curve which is constructed from large samples. 
 We present the rise time and 
 the decline rate in $B$-band from \cite{Ganeshalingam2011} 
 compared to those of SNe 2009ig, 2011fe, 2012cg, 2013dy and 2012ht
 in Figure \ref{rtc}.
 Underluminous objects which are spectroscopically 
 classified as SN 1991bg-like 
 \citep{Filippenko1992} have relatively short rise times 
 of 13-15 days 
 \citep{Modjaz2001,Taubenberger2008, Ganeshalingam2011}. 
 Figure \ref{rtc} shows that the rise time of SN 2012ht is significantly 
 larger than many of these underluminous events, but instead similar to 
 normal and overluminous SNe. 
 This implies that the transitional objects 
 can originate from the same explosion mechanism, progenitor 
 and/or population, with normal and overluminous events.

  \subsection{Progenitors and Explosions}

In this section we discuss implications obtained by our 
 data for a progenitor system and explosion mechanism of SN 2012ht. 
 The shock luminosity is predicted to be seen at early phase
 in several models \citep{Kasen2010,Piro2010,Piro2013,Rabinak2011}. 
 A shock luminosity could be determined by a progenitor radius, 
 an ejected mass, and opacity \citep{Piro2010,Rabinak2011}. 
 We assume that the ejecta mass and opacity of SN 2012ht are not significantly 
 different from those of SN 2011fe.
 Additionally, 
 shock luminosity exhibits the power-law decay with time evolution. 
 \cite{Bloom2012} gave a constraint on a progenitor 
 radius of SN 2011fe to be $\lsim$ 
 0.02~$R_{\odot}$, using an analytic model \citep{Kasen2010,Piro2010,Rabinak2011} 
 of the emission at $t=0.2$. 
 To estimate the progenitor radius of SN 2012ht, we scaled the parameters
 from SN 2011fe \citep{Bloom2012} in the three cases from \citet{Kasen2010,
 Piro2010,Rabinak2011}. 
 We calculate the putative shock luminosity of SN 2011fe at $t=1.8$ day 
 using the radius \citep[$\sim0.02~R_{\odot}$;][]{Bloom2012} in 
 each model \citep{Kasen2010,Piro2010,Rabinak2011}. 
 The radius of SN012ht is estimated by scaling this luminosity to 
 that of SN 2012ht. We give an upper limit on the radius as 
 1.5-2.7$R_{\odot}$, which indicates that the progenitor was 
 a (relatively) compact object.

\cite{Kasen2010} 
  suggested that a strong emission may come from the interaction 
 of the ejecta with its companion star. For a 2M$_{\odot}$ red giant (RG) 
companion star, the shock-induced emission reaches to $5.2\times10^{41}$ erg s$^{-1}$
at $t\lsim5$ day
even for the viewing angle opposite to the interacting 
direction \citep{Kasen2010}.
Since SN 2012ht is detected at $1.8\times10^{41}$ erg s$^{-1}$ at $t=1.8$ day,
our observations exclude the shock-induced emission by
the interaction with a RG companion.
It should be noted, however, that we do not reject the 
possibility 
that the RG might have already evolved at the time 
of the 
explosion through the spin-down scenario \citep{Justham2011}.


   If there is a variety in the $^{56}$Ni mass distribution within the outer 
 layers of SNe Ia, their rising parts are expected to exhibit some variations
 \citep{Piro2012b, Piro2013}.
 Our results indicate that the rising parts of SNe 2012ht, 2011fe
 and 2009ig are similar even including transitional SN~2012ht 
 (see the left panel of Figure \ref{rt}). 
 This finding indicates that the distribution of radioactive $^{56}$Ni in the outermost layer is similar among these three SNe Ia \citep[see also][]{Piro2012b, Piro2013}. 
 Various models predict different distribution of $^{56}$Ni near the 
 surface \citep[e.g., ][]{Gamezo2005, Sim2010, Seitenzahl2013}, and the `uniformity' we find indicates that 
 these SNe Ia, including transitional and normal SNe Ia, may share a 
 common explosion mechanism.

   \section{Conclusion}

   We successfully performed prompt observations of 
  SN Ia 2012ht. Our data show that SN 2012ht are very similar to 
  a transitional SN 2004eo. 
  The explosion date and the rise time are estimated as 
  MJD $56277.98 \pm 0.13$ and $17.6\pm 0.5$ days, respectively. SN 2012ht is the fifth 
  SN Ia for which such an accurate and strong 
  constraint on the rise time has been given directly by observational data.
  This rise time is comparable to that of normal or overluminous SNe Ia, 
 but much longer than that of underluminous SN 1991bg-like objects.
  This indicates that a common explosion mechanism could be the origin of normal and
 transitional SNe Ia. 

 

 \acknowledgements
 We would like to thank Mohan Ganeshalingam for data of rise time versus 
 decline rate of light curves. 
 This research has been supported in part in part by Optical \& Near-infrared
 Astronomy Inter-University Cooperation Program (OISTER) and by the Grant-in-Aid for 
 Scientific Research from JSPS (23340048,24740117,23244030) and 
 MEXT (25103515,24103003). Y.M. and R.I. have been supported by the
 JSPS Research Fellowship for Young Scientists.


  \newpage



 
 \newpage

  \pagebreak
  \newpage
 
  \newpage

  \newpage


   \clearpage


\end{document}